\documentclass[a4paper,twoside]{article}
\newcommand{\affil}[1]{$^{\rm #1}$}
\usepackage[authoryear]{natbib}
\bibpunct{(}{)}{;}{a}{}{,}
\usepackage{graphicx}
\date{} 
\title{\large\bf\flushleft Stellar Over-densities in the Outer Halo of the Milky Way}
\author{\parbox{\textwidth}{\flushleft
\vspace{-0.5cm}
{\it Stefan C.\ Keller\affil{A}\affil{B}}\\
\vspace{0.4cm}
{\small \affil{A}\,Research School of Astronomy and Astrophysics, Australian National University, \\Mt. Stromlo Observatory, Cotter Rd. Weston ACT 2611 Australia.}\\
{\small \affil{B}\,Email: stefan@mso.anu.edu.au}}}
\begin{document}
\twocolumn[
\begin{changemargin}{.8cm}{.5cm}
\begin{minipage}{.9\textwidth}
\vspace{-1cm}
\maketitle
%
%
\small{\bf Abstract: } This study presents a tomographic survey of a subset of the outer halo (10-40 kpc) drawn from the Sloan Digital Sky Survey Data Release 6. Halo substructure on spatial scales of $>3$ degrees is revealed as an excess in the local density of sub-giant stars. With an appropriate assumption of a model stellar isochrone it is possible for us to then derive distances to the sub-giant population. We describe three new candidate halo substructures; the 160- and 180-degree over-densities (at distances of 17 and 19 kpc respectively and radii of 1.3 and 1.5 kpc respectively) and an extended feature at 28 kpc that covers at least 162 square degrees, the Virgo Equatorial Stream. In addition, we recover the Sagittarius dwarf galaxy (Sgr) leading arm material and the Virgo Over-density.

The derived distances, together with the number of sub-giant stars associated with each substructure, enables us to derive the integrated luminosity for the features. The tenuous, low surface brightness of the features strongly suggests an origin from the tidal disruption of an accreted galaxy or galaxies. Given the dominance of the tidal debris of Sgr in this region of the sky we investigate if our observations can be accommodated by tidal disruption models for Sgr. The clear discordance between observations and model predictions for known Sgr features means it is difficult to tell unambiguously if the new substructures are related to Sgr or not. Radial velocities in the stellar over-densities will be critical in establishing their origins.


\medskip
\medskip
\end{minipage}
\end{changemargin}
]
\small

\section{Introduction}
How does a galaxy such as our Milky Way form? This remains a pressing question for modern astrophysics. The seminal work of \citet{ELS} proposed that galaxy formation occurred in the wake of a monolithic collapse of an isolated proto-galactic cloud. \citet{SZ} challenged this picture of isolated formation by proposing instead that the halo was assembled by the accretion of numerous small entities over an extended period.

Recent observational evidence suggests both formation scenarios have a role to play. \citet{Carollo07} presents evidence that the inner halo (that of galactocentric radius $<15$kpc) is dominated by highly eccentric, prograde orbits and a metallicity of [Fe/H] $\sim -1.6$ due to in-situ formation, whereas the outer halo shows a more uniform distribution of eccentricities, includes highly retrograde orbits, and lower mean metallicity ([Fe/H] $\sim -2.2$) that, they argue, has an accretion origin. The \citet{Kinman07} study of the local halo, similarly identifies a population that possesses retrograde rotation and streaming motions (i.e.\ low velocity dispersion) and another population that exhibits negligible rotation and less indication of streaming. \citet{Kinman07} show that the horizontal branch morphologies of these two halo components are consistent with those seen in the young and old globular cluster systems respectively. This reinforces the conclusion of \citet{Zinn85} that the inner halo is consistent with formation from rapid collapse and that the outer halo was formed from ongoing accretion of dwarf galaxies.

Simulations of galaxy formation in the context of the prevailing $\Lambda$CDM cosmology predict that the majority of stars in the outer halo formed in progenitor dwarf galaxies that were subsequently accreted (see for example, \citet{Freeman02}). \citet{Bullock05} show that spatially coherent substructures formed of tidal debris should survive in the outer halo for several Gyrs and that their kinematic signatures (in velocity space) should remain detectable well after spatial substructure is no longer evident \citep{Helmi03}. On the other hand, a halo that formed from in-situ star formation would possess, at the current epoch many dynamical times later, negligible substructure. \citet{Bell07} compare the level of substructure seen in the SDSS data with cosmological simulations (over galactocentric distances 1-40 kpc) and find that the Milky Way's halo is consistent with assembly from dwarf galaxies.

The most prominent contribution to outer halo substructure results from the on-going disruption of the Sagittarius Dwarf Spheroidal galaxy \citep[Sgr, ][]{Ibata94}. The leading and trailing arms of Sgr are seen to encircle the sky \citep{Majewski03, Newberg02, Newberg07, fieldofstreams, Keller08}. The extensive debris of Sgr offer an important probe of the mass and shape of the Galaxy \citep{Ibata01, Helmi01, Helmi04, Johnston05, Law05, Fellhauer06}.

Another outer halo substructure, the Virgo Over-density (VOD) was found as an over-density in RR Lyrae variables \citep[RRLs, ][]{Vivas01, Vivas04, VivasZinn06, Keller08} and in SDSS as a diffuse structure of main-sequence stars \citep{Juric08, Newberg02, Newberg07} and blue horizontal branch stars \citep[BHBs, ][]{Ivezic05}. Radial velocity measurements by \citet{Duffau06}, \citet{Newberg07}, \citet{Vivas08}, and \citet{Prior09} reveal that a subset of associated RRLs share a common spatial velocity. This moving group is termed the Virgo Stellar Stream (VSS) to distinguish it from the general stellar over-density and potentially other velocity groups that may spatially coincide.

The study of \citet{Duffau06} explored the spatial extent of the VOD by considering the excess of the observed luminosity function over an off-source control field. This study discerned a size for the VOD of at least 106 square degrees. A similar technique is implemented by \citet{Prior09} to derive a spatial extent for the VOD of $\sim760$ square degrees. The study by \citet{Juric08} utilized the photometric parallaxes of F-type main-sequence stars to describe a large diffuse over-density covering $\sim 1000$ square degrees and spanning a range of distance from 6-20 kpc. \citet{Vivas08} find a small concentration of stars with similar radial velocities to the VSS reaching to 12 kpc from a previous detection at 19 kpc.

\citet{MartinezDelgado04,MartinezDelgado07} suggest that the VOD is the result of the confluence of the leading and trailing arms of Sgr. However, as pointed out by \citet{Newberg07} and \citet{Vivas08}, models of the Sgr debris stream predict highly negative radial velocities contrary to the $+$100-130 kms$^{-1}$ seen in the VSS. Extant models also fail to match the observed density enhancement of the VOD. \citet{Newberg07} and \citet{Vivas08} conclude that the VOD is most likely tidal debris from a separate and distinct accretion event. Recently, \citet{Casetti-Dinescu09} have presented a preliminary orbit for the VSS based on one RRL that places the system near pericentre on an eccentric and highly destructive orbit.

In light of the considerable uncertainty regarding the apparent size, distance and origin of the VOD we have presented a series of new observational studies targeting this region. \citet{Keller08} has examined the spatial distribution of RRL candidates in the VOD region and finds in addition to the over-density seen by the QUEST team at 17 kpc, a second region of over-density 15 degrees to the south-east and at a distance of 20 kpc. The study of \citet{Prior09} examines the kinematics of a sample of RRL candidates from \citeauthor{Keller99} et al. (2008). In contrast to previous studies, this study finds RRLs that possess radial velocities consistent with membership of Sgr trailing arm debris. They conclude that a significant contribution from Sgr can not be ruled out at the present time. \citet{Keller09_TNO} utilizes the luminosity function excess over that expected from the Besancon galaxy model \citep{Robin03} to map the extent of the VOD. This study finds that the region consists of three spatially distinct over-densities that extend from declinations $+$2$^{\circ}$ to $-$15$^{\circ}$.

In the present study, we utilize the sub-giant population (those stars between the main-sequence turn-off and the base of the red giant branch) to perform tomography of the outer halo in order to describe the VOD in more detail and to look for additional substructure. In Section \ref{section:subgiant_tomography} we present our method for determining the significance of, and distance to, halo substructure. Section \ref{section:results} presents our results on the extent and luminosity of a series of halo sub-structures with comparison, where possible to existing literature. Finally, in Section \ref{section:origins} we discuss our map of substructure in relation to the tidal debris of Sgr.

\section{Sub-giant tomography}
\label{section:subgiant_tomography}

Our technique is to take an identifiable stellar population and use this population as a tracer of spatial density by translating apparent magnitude to distance for the population. As discussed previously, this technique has utililsed RRLs, BHBs, M giants and main-sequence (MS) turn-off stars. Here we utilise the sub-giant population. 

There are a number of advantages to the use of sub-giants.Firstly, sub-giants are a numerous population with which to trace substructure. Commonly used distance indicators like RRLs, BHBs and M giants while presenting a population free from high contamination, are too sparse (typical values are 0.1 per square degree for M-giants to 10 per square degree in the case of BHBs) to present substantial contrast over their surrounds to probe spatial structure on small angular scales. 

The MS population presents a much more numerous population ($\sim 7500$ per square degree; see \citet{Juric08}). Plentiful as they are, deriving distances to MS stars is fraught with considerable difficulty. As discussed in \citet{Juric08} conversion from the apparent magnitude to the distance of a MS star requires a well described photometric parallax relation. The technique requires highly accurate colours in order to precisely derive an absolute magnitude. As \citet[their figure 7]{Juric08} shows, distances to stars at the MS turn-off possess a fractional distance uncertainty of between $\pm15-\sim65$\%.  This is a result of the extremely steep absolute magnitude - colour relation inherent in the photometric parallax method. In addition, considerable systematics enter with the choice of a photometric parallax relation (up to $\pm70$\% in distance for the bluest halo stars).

Sub-giants, on the other hand, are less numerous but are significantly cooler stars than those at the MS turn-off. Furthermore, these objects result from a evolutionary phase of limited duration, as opposed to the MS. We can therefore, more precisely derive their distance by relating their apparent brightness to that predicted by stellar evolution isochrones. 

Sub-giants do, however, suffer from an ambiguity since they are selected from a window of the colour-magnitude diagram (CMD) that is inhabited by core helium burning stars on the red clump and horizontal branch, and lower MS stars. This has the potential to lead to three peaks in measured stellar density as a function of magnitude from a population of set distance as each of three populations are encountered at intervals of approximately 3 magnitudes. If populations of similar colours to the sub-giant branch are present  they will lead to erroneous conclusions as to the distance of such over-densities. A population who's horizontal branch and/or red clump appears in the sub-giant window will be assigned a distance modulus 3 magnitudes, or 4 times too close. We bear this caveat in mind in the discussion to follow.

The region we have chosen to investigate in detail is the ``Field of Streams" described by \citet{fieldofstreams}. This region extends from declinations $+34^{\circ}$ to $-4^{\circ}$ and covers Right Ascension 120$^\circ$ to 220$^\circ$. The area encloses the northern leading arm of the Sgr Stream and the VOD. We examine the significance of sub-giant excess over the nominal field population in square `pixels' drawn from the photometry of SDSS DR6 \citep{Adelman-McCarthy08}. 

\subsection{The determination of the significance of sub-giant excess}
\label{section:method}
\begin{figure}[h]
\begin{center}
\includegraphics[height=90mm]{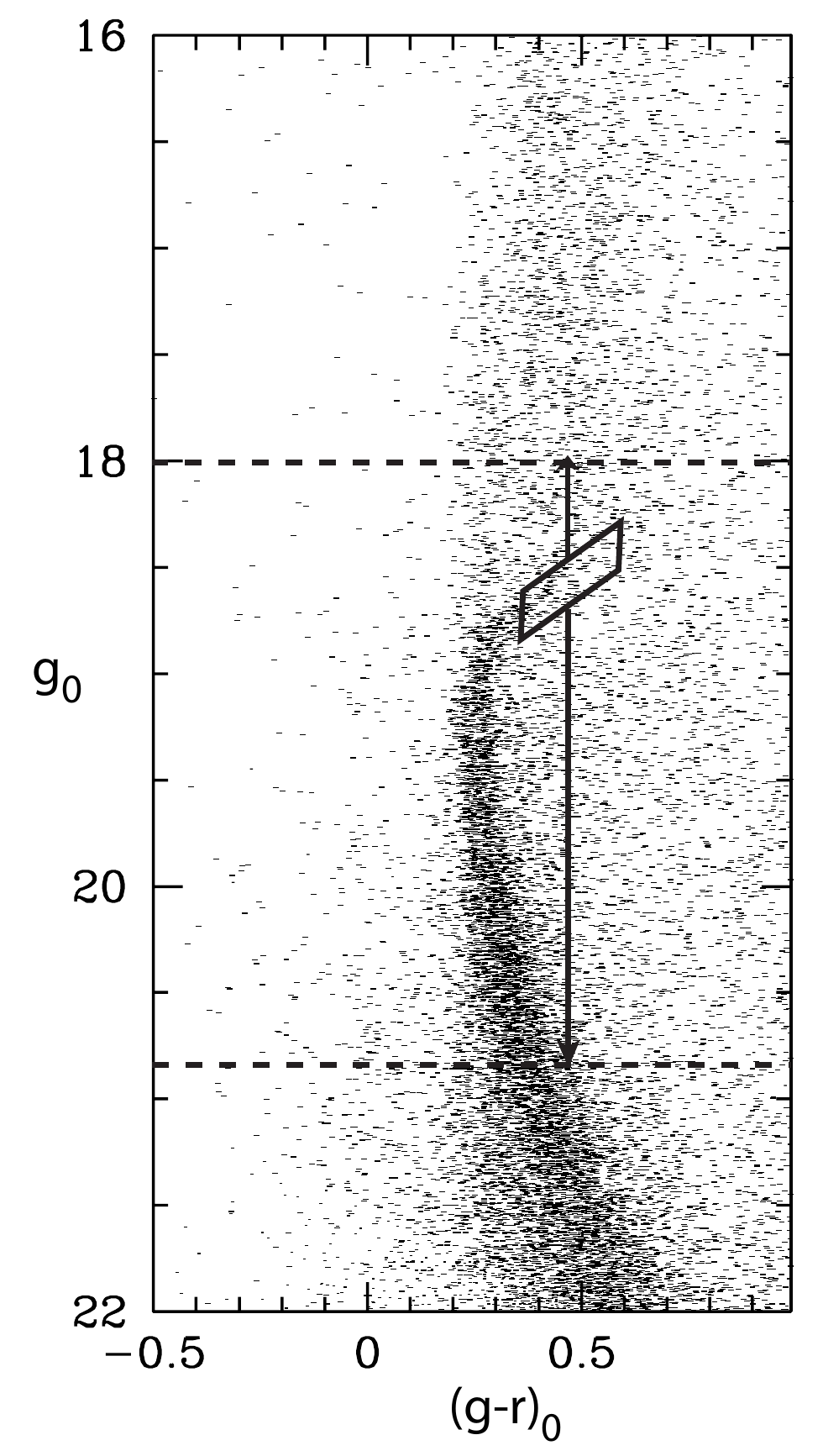}
\caption{An illustration of our selection of sub-giant stars from the field containing the globular cluster M3. The selection box (0.35$<(g-r)_{0}<0.60$ and 0.2 mag. in height) shown is placed between 18.0$<g_0<$21.7 mag.\ at 0.1 mag.\ steps and the number of stars so selected recorded as a function of magnitude.}\label{figure:M3example}
\end{center}
\end{figure}

\begin{figure}[h]
\begin{center}
\includegraphics[scale=0.55, angle=0]{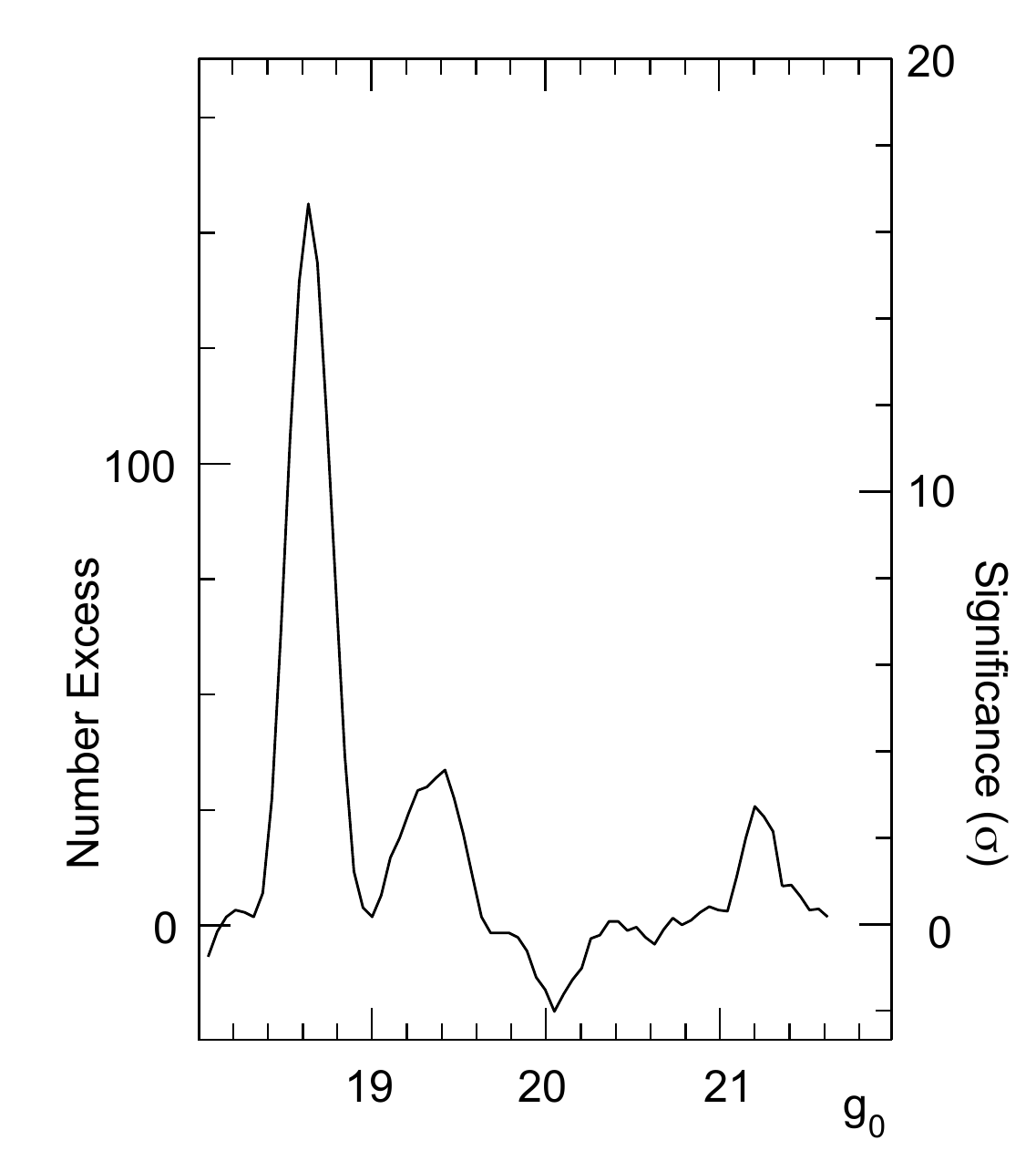}
\caption{A plot of the number of excess sub-giant stars as a function of magnitude resulting from the selection process described in Figure \ref{figure:M3example}. The prominent peak seen corresponds to selection of the M3 sub-giant branch. Using Monte-Carlo simulations we are able to quantify the excess of sub-giant stars in terms of statistical significance (right-hand axis, see text for details).}\label{figure:signifexample}
\end{center}
\end{figure}

Our procedure of determining the significance of sub-giant excess at a position on the sky is as follows. We determine the number of objects within a window on the colour-magnitude diagram (CMD) that follows the locus of the globular cluster M3 sub-giant population between $0.35<(g-r)_{0}<0.60$ and is 0.2 magnitudes in height as shown in Figure \ref{figure:M3example}. The box is then placed at $g_0$ magnitudes from 18.0 to 21.7 in 0.1 magnitude steps. Fainter than $g_0$=21.7 photometric uncertainties  scatter halo turn-off stars into our sub-giant selection window. This reduces the contrast of possible superimposed substructures. The bright limit of our magnitude range is set to avoid contribution for the thin and thick disk main-sequence stars to our sub-giant sample. This defines the effective distance range for the present study of 10-40 kpc. As seen from simulations of \citet{Robin03} (and furthermore, by observations by \citet{Conn08}), halo stars are dominant for  $g_0>18.0$. We then subtract a low-order polynomial fit to the number of sub-giant stars as a function of $g_0$ magnitude. This leaves us with departures from the general background as shown in Figure \ref{figure:signifexample}.

The spatial pixel size is 3$^\circ \times 3^\circ$. Such pixels contain typically 200 stars within the colour selection window at each magnitude step (hence a Poisson uncertainty of $<7$\% per magnitude step). Making the pixel size smaller introduces excessive shot noise and making the pixel larger removes spatial information. The relatively coarse grid size leaves us sensitive to extended stream structures but unable to resolve ``fine-grain'' (<3$^\circ$ wide) structures such as the Orphan Stream \citep{Belokurov07} or the streams of \citet{Grillmair09}.

We then consider the region 180$^{\circ}<$RA$<190^{\circ}$ and $24^{\circ}<$Dec$<34^{\circ}$ that, as evidenced by the CMD of the region, does not contain significant over-density. As can be seen from \citet{fieldofstreams} the northern branch of the bifurcated Sgr leading arm does pass through this area. However, it is sufficiently distant that it does not significantly enter the distance range probed by our sample. The quoted distance to this Sgr debris is, we contend, underestimated due to an assumed high metallicity as discussed below. From this large area we draw multiple samples that match the sample size typical of our 3$^\circ \times 3^\circ$ fields. This enables us to quantify the stochastic variation inherent in Figure \ref{figure:signifexample}. We can then transform the measured excess of sub-giants over the smooth field to a significance of sub-giant excess as a function of magnitude. 

By recourse to theoretical isochrones, we can relate the $g_0$ magnitude at which the excess is observed to a distance. We have assumed an isochrone of 12 Gyrs and [Fe/H]=-1.7 \citep{Girardi04} that is typical of the population expected from a dwarf spheroidal galaxy \citep{Mateo98} and is motivated by our previous metallicity determinations for RR Lyraes in the VOD \citep[average of -1.7dex, ][]{Prior09}. Furthermore, this is in line with the recent determination of the metallicity in the vicinity of the VOD by \citet{An09} who find [Fe/H]=-2.0$\pm0.5$dex. 

As shown in Figure \ref{figure:signifexample} for M3, we derive a distance modulus of 15.10$\pm0.02$. This is in good agreement with the distance modulus of 15.09 from \citet{Harris96}. The distance modulus derived is only slightly dependent on the choice of age (e.g.\ a choice of 10 Gyrs results in a distance modulus of 15.12) but moderately dependent on the choice of metallicity, for example, assuming [Fe/H]=-1.3 results in a distance modulus of 14.94; [Fe/H]=-0.7 results in a distance modulus of 14.52.

\section{Sub-structure revealed}
\label{section:results}

The significance of sub-giant excess as a function of distance was computed in each $3^{\circ} \times 3^{\circ} \times 0.6kpc$ `voxel' in our survey area. The result is a data cube of over-density significance. For the purposes of illustration we have projected the data cube along the shortest axis (Declination). To form this projection we take the sum of the significance along the 12 pixels of the Declination axis.

\begin{figure*}[t]
\begin{center}
\includegraphics[scale=0.55, angle=0]{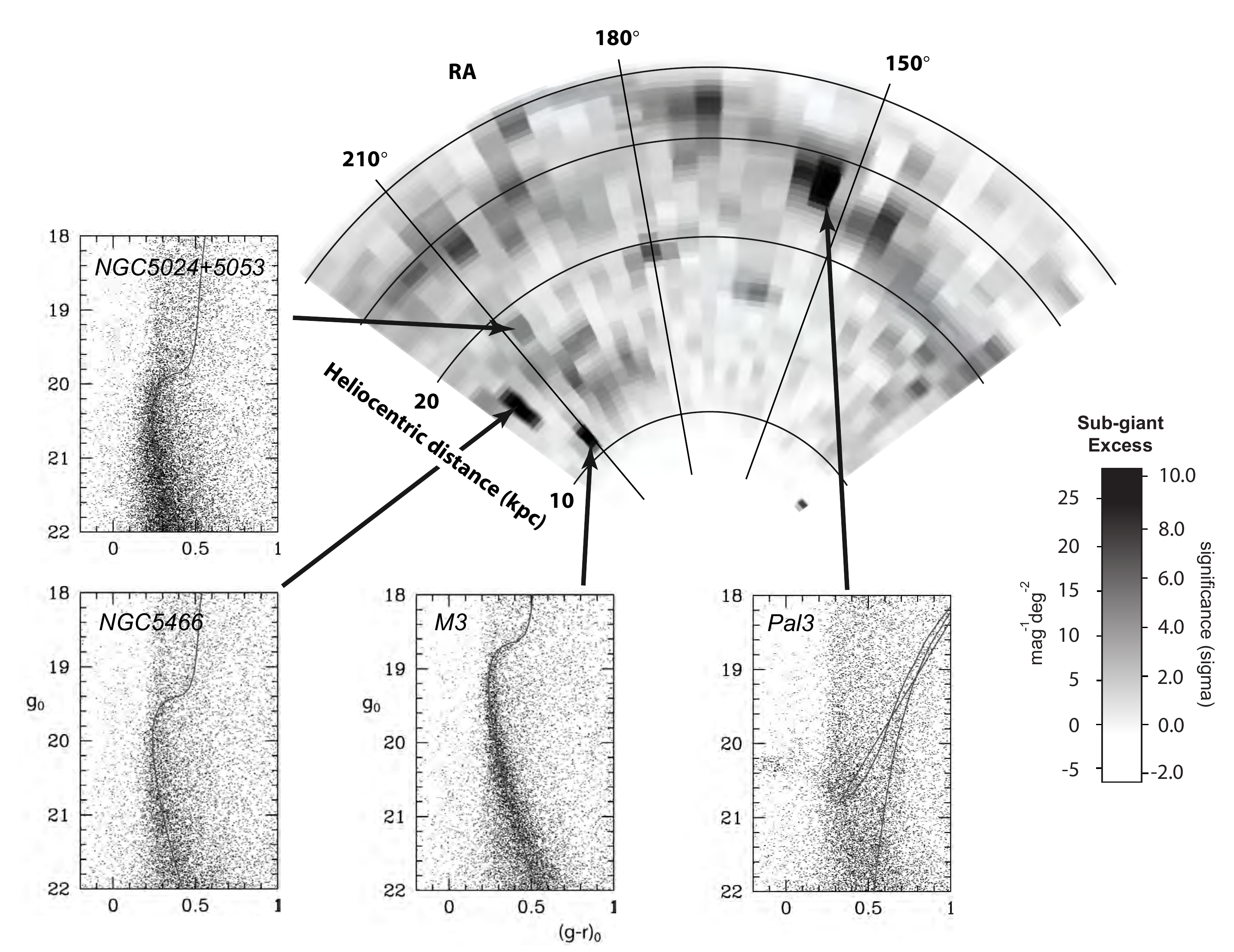}
\caption{The excess of sub-giant stars seen in our survey volume. In this figure the excess data cube is projected along the declination axis. It shows the prominent leading arm of Sgr tidal debris that enters the survey volume at RA=120$^{\circ}$ and a heliocentric distance of 20 kpc and exits at RA$\sim190$$^{\circ}$ at 40 kpc, the distance limit of our survey. The four inset panels show the colour-magnitude diagrams of 3$^{\circ} \times 3^{\circ}$ fields that contain prominent globular clusters. The isochrones shown for M3 and NGC 5466, 5024 and 5053 are due to \citet{Girardi04} (age=12Gyrs; [Fe/H]=-1.7) placed at the distance modulus derived from the peak of sub-giant significance from our technique discussed in Section \ref{section:method}. The isochrone for Pal3 is shown at a distance modulus of four times that measured by our technique to account for the red clump population (see Section \ref{section:method} for details).}\label{figure:gcs}
\end{center}
\end{figure*}

A number of sources were seen that correspond to single spatial (i.e. RA,Dec) pixels and confined to a narrow range in distance. These are due to known globular clusters in the field. In Figure \ref{figure:gcs} we show the the CMDs for four fields containing prominent globular clusters. The distances to three of the globular clusters are in good agreement with those in the literature \citep[10.4, 14.4, 16.5kpc compared with 10.4, 15.9, 17.1kpc from][for M3, NGC5466 and NGC5024+5053 respectively]{Harris96}. The CMD for Pal3 appears at a distance of 24 kpc, a factor of four closer than its literature value \citep[92.7kpc,][]{Harris96} due to the inclusion of its red clump in the sub-giant selection window.

\begin{figure*}[h]
\begin{center}
\includegraphics[scale=0.50, angle=0]{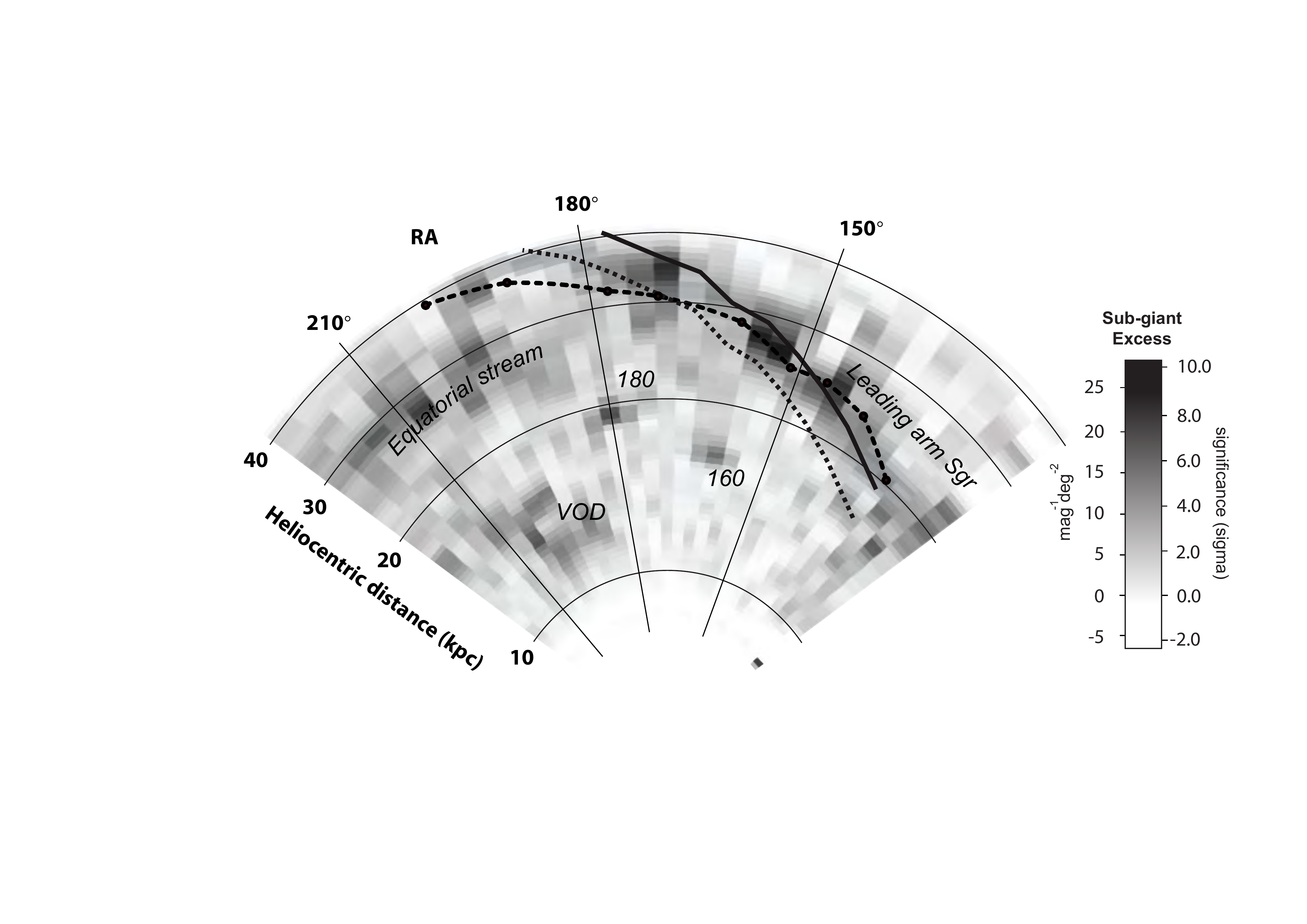}
\caption{As in Figure \ref{figure:gcs}, the excess of sub-giant stars seen in our survey volume projected along the declination axis. Spatial pixels associated with globular clusters are ignored in creating this figure (see text for details).  Overlaid is the locus of the Sgr leading arm as described by \citet[][dashed line with points]{Newberg07}. The dotted line shows the locus due to \citet{fieldofstreams} and the solid line is the same result assuming a [Fe/H]=-1.1 as found by \citet[][see text for details]{Chou07}. The Virgo over-density is seen over RA=184$^{\circ}$-210$^{\circ}$ and distances of 12-18 kpc. Three new candidate over-densities are seen, the Virgo Equatorial Stream and the 160$^{\circ}$ \& 180$^{\circ}$ features. }\label{figure:nogcs}
\end{center}
\end{figure*}

In the following analysis of the data-cube collapsed along the Dec axis we have excluded the RA, Dec pixels that  contain galactic globular clusters (this excises 1.7\% of the survey volume and hence is unlikely to modify our conclusions).  Figure \ref{figure:nogcs} is dominated by the Sgr leading arm. We utilise this feature as a further useful check of our methodology. The leading arm enters our survey volume at RA=$122^\circ$ and at a distance of 20 kpc and extends to our survey limit of 40 kpc at RA$\sim 195^\circ$. We can compare our findings with the locii of the leading arm presented in \citet{fieldofstreams} and \citet{Newberg07}.

The distances in the study of \citet{fieldofstreams} are calculated from the relative offset between the colour-magnitude diagram (CMD) of the debris and that of the main body of Sgr. Since the most prominent feature of the CMD is the subgiant branch, the technique of \citet{fieldofstreams} is analogus to that used here. However, the main body of Sgr has a mean metallicity of $[$Fe/H$]$=-0.4 \citep[][with a considerable low metallicity tail to Fe/H $\sim-1.6$]{Smecker-Hane02, Monaco05, Chou07} and the debris has mean [Fe/H]=-1.1 \citep[but with reduced spread in Fe/H relative to the core,][]{Chou07}. Isochrones of \citet{Girardi04} show that this difference in mean metallicity leads to potentially a 0.4 mag.\ under-estimation in the derived distance modulus of the Sgr leading arm material. If we apply this shift to the \citet{fieldofstreams} distances (the dotted line in Figure \ref{figure:nogcs}) we recover the solid line seen in Figure \ref{figure:nogcs}.
 
\citet{Newberg07} distances are derived using photometric parallaxes of main-sequence F-type stars with a distance calibration from Sgr BHBs. As above, systematic changes in metallicity along the leading arm may induce systematic divergence in derived distance. Nonetheless, as can be seen from Figure \ref{figure:nogcs}, the assumption of low metallicity produces a degree of concordance between the two previous determinations and the present study.

\subsection{Discussion of substructures}
\label{section:discussion_of_substructure}

In addition to the Sgr leading arm, four spatially distinct regions of over-density are apparent in Figure \ref{figure:nogcs}. We identify these as the Virgo Over-density (VOD) that has been known from previous studies discussed above and new candidate stellar over-densities I term the 160$^\circ$ and 180$^\circ$ features, and the Virgo Equatorial Stream (VES). 

The first issue we must consider is if these features could result from a background population of red clump stars in our sub-giant colour selection window, due, for instance, to the Sgr stream.  In order to produce the Virgo Equatorial Stream, 160$^{\circ}$ and 180$^{\circ}$ over-densities the Sgr stream red clump stars would have to lie at $\sim110$, $\sim65$ and $\sim75$kpc respectively. As evident from numerous studies of Sgr (see above discussion, and shown here in Figure \ref{figure:nogcs}) the leading arm is a factor of two closer at these positions and so can not account for the observed over-density via its population of red clump stars. In the case of the VOD, the distance to its associated RRL over-density is in concordance the distances derived here. 

Similarly, to bring about the observed over-densities be means of lower main-sequence stars falling in the sub-giant colour selection window would require distances of approximately 7, 4, and 5kpc for the Virgo Equatorial Stream, 160$^{\circ}$ and 180$^{\circ}$ over-densities. There are no indications from the distribution of RRLs at these distances that such over-densities do exist \citep{Miceli08}. 

Secondly, the features do appear to be artefacts of sampling to a coarse spatial grid. They are not expunged by a different choice of spatial pixel centroids. We have experimented with sub-pixel spatial shifts of the grid and recover the over-densities at similar significance. 

\begin{figure*}[h]
\begin{center}
\includegraphics[scale=0.4, angle=0]{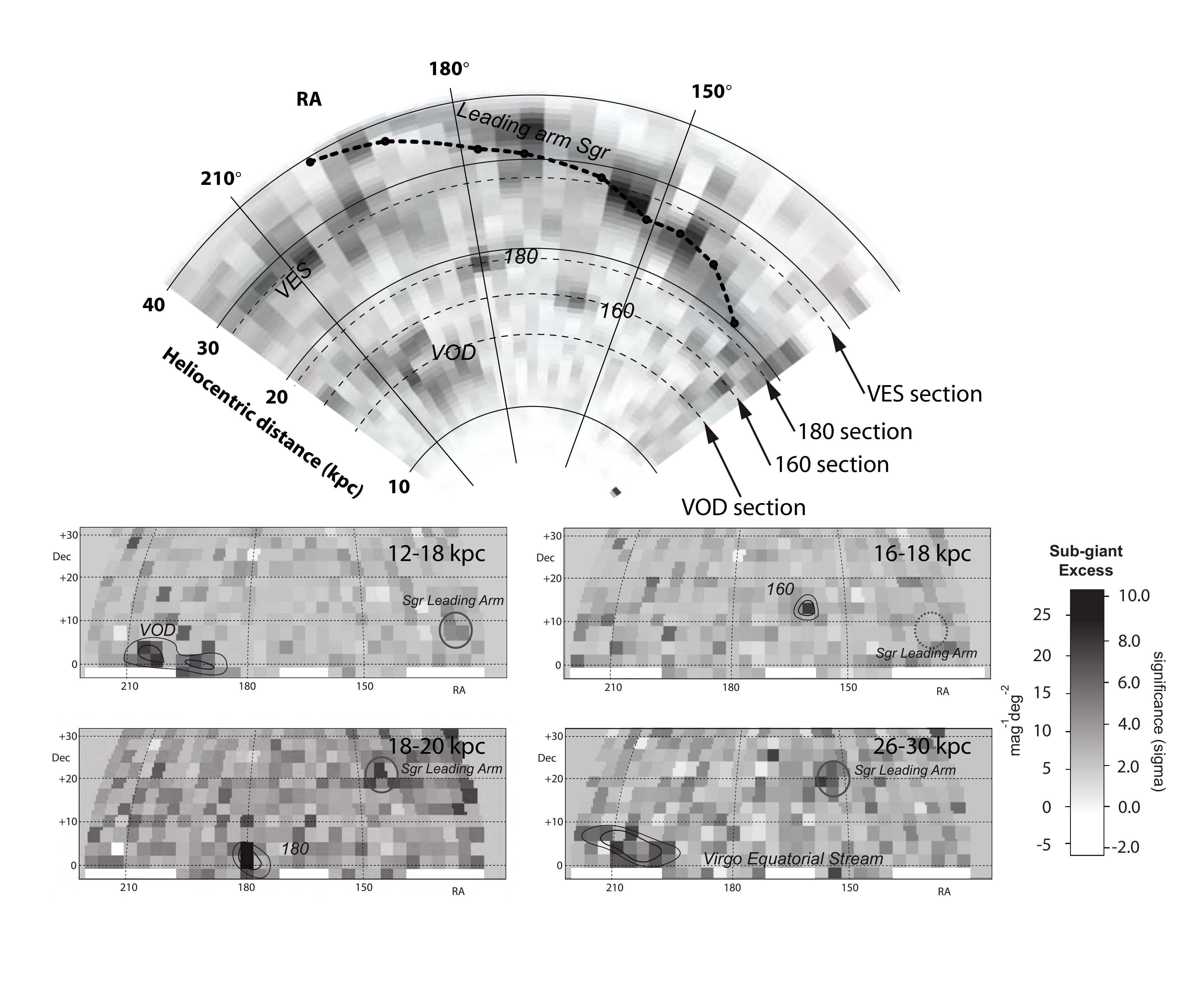}
\caption{{\bf{Top}}: Dashed lines show the distance at which sectional slices through the significance data cube are made to reveal the spatial orientation and morphology of the identified substructures. 
{\bf{Bottom}}: RA, Dec slices through the excess data cube at the mean distances of each sub-structure. The distance range averaged in each panel is given in the top right. The contours around each sub-structure shows the extent of the structure. The inner contour contains 50\% of the sub-giant over-density, and the outer contour contains 75\% of the over-density (see Section \ref{section:discussion_of_substructure} for details). Slices through the Sgr leading arm can be seen in the projected sky significance maps for the VOD, the more distant 160$^{\circ}$ feature and the Virgo Equatorial Stream.}\label{figure:pos}
\end{center}
\end{figure*}

The region examined in the present study has been the focus of a number of previous studies targeting halo substructure. In the following section we discuss our results in the context of these previous studies where possible. To define the extent of each feature we implement a kernel discriminant analysis. The technique is a non-parametric smoothing of the spatial distribution by an optimally selected multivariate kernel. The result is a probability density function for the data that retains maximum information content afforded by the data (i.e.\ not over/under smoothed). Such probability density function were constructed (using the {\tt{ks}} package \citep{Duong07} in {\tt{R}}) for all four features. Contours that contain 75\% and 50\% of feature data (drawn from a broad rectangular volume containing the feature) are shown projected in two dimensions in Figure \ref{figure:pos}. The corresponding areas of the features are given in Table \ref{table:mvs}. 

We draw the reader's attention to the fact that these are sectional slices through the data cube and so as such only show those features visible at a specific distance. For this reason the Sgr leading arm does not appear as a contiguous element but as a 6-9 degree wide section in these panels. The Monoceros ring \citep{Newberg02} (or warp, \citet{Conn08}) falls outside the field under consideration here at lower galactic latitudes off the right of Figure \ref{figure:nogcs}.

\begin{figure}[h]
\begin{center}
\includegraphics[scale=1, angle=0]{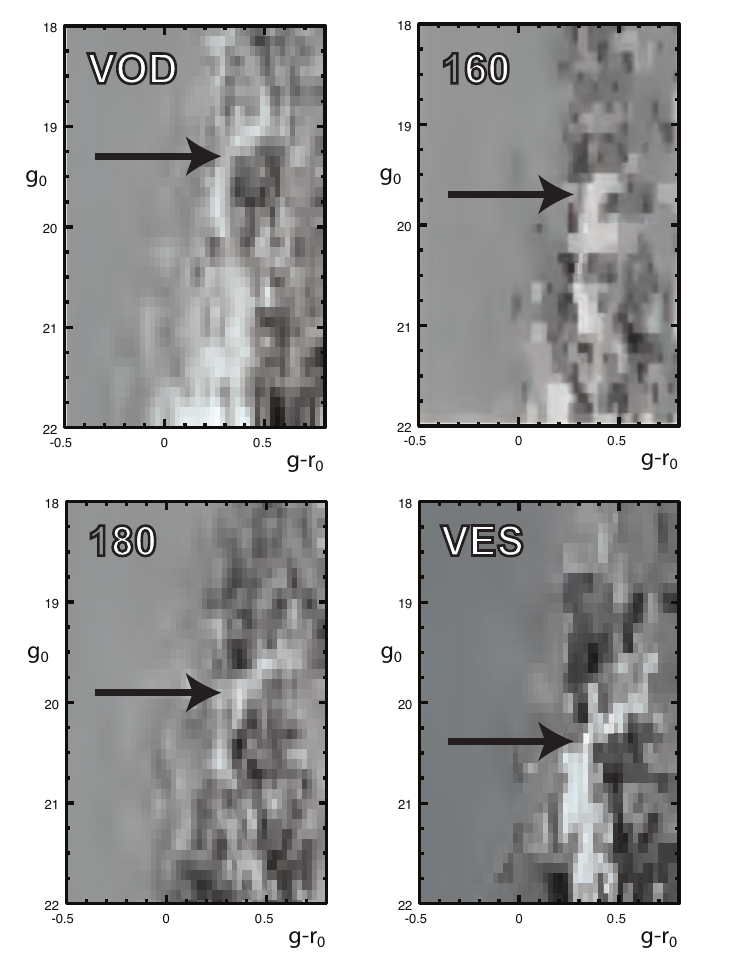}
\caption{Hess diagrams of the population of the sub-structures discussed in the present study. We have subtracted from the on-source field a population taken from an adjacent off-source field, appropriately scaled to match the total number of stars in the on-source field. Lighter greyscale indicates population excess. The arrows indicate the approximate magnitude of the main-sequence turn-off in each case.}\label{figure:cmds}
\end{center}
\end{figure}

In Figure \ref{figure:cmds} we show the Hess diagrams for each sub-structure population. These diagrams are derived by taking the on-source population CMD density and subtracting an adjacent off-source CMD density, appropriately scaled to match the number of stars in the on-source field. Here, $g$ and $r$ magnitudes have been dereddened according to the reddening maps of \citet{Schlegel98} together with a colour-specific reddening index of E($g$$-$$r$) = 1.042E($B$$-$$V$) and $A_g $/E($B$$-$$V$) = 3.793. The Hess diagrams reveal low density main-sequence turnoff features for each sub-structure. For example, Figure \ref{figure:cmds} (top left) shows the $g_0 \sim 19$ turnoff for the VOD. In this example, the more distant Sgr leading arm material can be seen in the sub-giants at $g_0 \sim 21$.

\subsubsection{The Virgo over-density}
\label{section:VOD}

The VOD is seen to be a diffuse structure centred at RA=194.6$^{\circ}$, Dec=+1.8$^{\circ}$ covering 144 square degrees and distances from $\sim12$-$18$ kpc. With a peak significance at a distance of 16 kpc, the projected dimensions of the VOD in RA is 5 kpc and 1.7 kpc in Dec. We note that \citet{Keller09_TNO, Prior09, Newberg07} (discussed below) have shown that the VOD extends outside the bounds of the SDSS survey to the south-west and hence the spatial extent derived here is a lower limit. Further more, the significance measured along the line of sight for the VOD is also underestimated due to the depth of the structure. This underestimation occurs because our method for determining significance requires the subtraction of a low order polynomial fit to the general background. Our simulations show that this has the effect of diminishing the significance of, but not erasing, features that exhibit a line of sight depth of $g>0.4$ mag.

The VOD was first detected as an over-density in RR Lyrae variables (RRLs) by \citet{Vivas01} and independently detected as an over-density of F-type main-sequence stars by \citet{Newberg02}. Both studies describe the centre of the VOD at RA=190$^{\circ}$, Dec=0$^{\circ}$ and a heliocentric distance of 18 kpc. From a 2.3$^{\circ}$-wide band centred on Dec=-1.18$^{\circ}$, \citet{VivasZinn06} find a peak density for the VOD at RA=186$^{\circ}$ and a distance of 17kpc. In the study of RRL candidates by \citet{Ivezic05} a distance of 18 kpc is derived for the VOD with a center at RA$\sim190$$^{\circ}$ in a 2.5$^{\circ}$-wide band centred on Dec=0$^{\circ}$. 

\citet{Duffau06} presents radial velocities for a sample of \citet{Vivas01} RRLs and \citet{Sirko04} BHBs.  They find six of the nine stars in the densest region of the VOD (centred at RA=186$^{\circ}$ and Dec=-1$^{\circ}$) share a common radial velocity of $+100$ kms$^{-1}$ with a velocity dispersion less than measurement uncertainties. This moving group was termed by \citeauthor{Duffau06} the Virgo Stellar Stream to differentiate it from the general region of over-density. Recently, \citet{Vivas08} finds members of the VOD that share this common radial velocity (i.e.\ the VSS) that extend as close as 12 kpc. By quantifying the excess in the luminosity function in VOD fields relative to offset fields, \citet{Duffau06} reports that the VOD covers an area in excess of 106 square degrees.

It was demonstrated by \citet{Newberg07} that the VOD extends beyond the southern edge of the SDSS survey. \citeauthor{Newberg07} shows significant over-density that appears a continuation of the VOD in one of the three SEGUE `outrigger' scans. \citet{Newberg07} examines SDSS spectroscopy in a 1.5$^{\circ}$ radius field centred on the peak of this over-density at RA=191$^{\circ}$ and Dec=-7.8$^{\circ}$. They find a marginally significant peak in the radial velocity distribution in common with the VSS. This suggests that the VSS extends across this area.

\citet{Juric08} presents a study of the density excess of main-sequence F-type stars towards the VOD. They present evidence that the VOD covers $\sim$1000 square degrees centred on RA=192$^{\circ}$ and Dec=+2$^{\circ}$ and has extensive line-of-sight depth (6-20 kpc). The study of \citet{Prior09} implements a technique similar to that of \citet{Duffau06} that utilizes the excess in the observed luminosity function. In this case, the excess is relative to Besancon galaxy model \citep{Robin03} predictions. This study estimates an areal extent of the VOD of 760 square degrees.  

\citet{Keller09_TNO} matches the technique of \citet{Prior09} over 1$^{\circ} \times 1^{\circ}$ fields from the SDSS DR6 \citep{Adelman-McCarthy08} and SEKBO \citep{Keller08} datasets. The SEKBO data enables an extension to the south-west that reveals that the region of the VOD encompasses three regions of over-density. \citet{Keller09_TNO} Feature $A$ is a linear feature approximately 3$^{\circ}$ wide and 14.5$^{\circ}$ long centred at (198$^{\circ}$, -10$^{\circ}$) with a distance of 20 kpc (hence linear size of 1$\times$5 kpc). Feature $A$ is not seen here as it does not extend into the SDSS sample. Feature $B$ at (192$^{\circ}$, -2$^{\circ}$) is 3$^{\circ}$ wide and 10$^{\circ}$ long, with a heliocentric distance of 17 kpc (linear size of 1$\times$3 kpc). It is coincident with, and hence identified as, the VOD. A third over-density, Feature $C$, is identified with the 180$^{\circ}$ feature of the present study (discussed below).

The extent of the VOD seen in the present study, and that of \citet{Keller09_TNO}, is substantially smaller than that of \citet{Juric08} or \citet{Prior09}. The study of \citet{Juric08} derives the spatial extent of the VOD from the density of stars within 0.2$<g-r<$0.3 and 20$<r<$21, so chosen to select main-sequence stars at a distance appropriate for the VOD. \citet{Newberg07} applies identical selection criteria, but for bracketing distances ($r$ magnitudes one magnitude brighter and fainter; see \citet{Newberg07} figures 5 and 6). The results of Newberg et al.\ show that the Sgr stream remains a significant contributor to the stellar density over 20$<r<21$. We propose that the northern portion of the stellar density seen the Juric et al.\ result (their figure 37) is due to the Sgr leading arm rather than the VOD, and that this leads to Juric et al.\ to greatly overestimate the spatial extent of the VOD. In addition, the \citet{Juric08} study utilizes photometric parallax to determine the distance to the target main-sequence stars. As discussed previously, the inherently large uncertainties associated with this technique lead to an {\it{over}}estimation of the depth of the VOD by \citeauthor{Juric08}. In the case of the \citet{Prior09} study, the spatial sampling proved insufficient to resolve the VOD region into the series of distinct components seen by \citet{Keller09_TNO}. 

The VOD possesses an apparent line-of-sight inclination on the sky such that the closest region at 12 kpc is to the eastern extremity (RA$\sim 184^\circ$) and the furthest region at 18 kpc is at the westerly extremity (RA=$210^\circ$). This is in line with the findings of \citet{Keller09_TNO} who, from distances to the RR Lyrae members, finds the westerly portion to be at a distance of 20 kpc from a central region (at RA=194$^{\circ}$) of 17 kpc. Since \citet{An09} finds no significant metallicity gradient across the VOD region, this is likely due to a true spatial gradient.

\subsubsection{180$^{\circ}$ over-density}

The 180$^{\circ}$ over-density (centred at RA=179.8$^{\circ}$, Dec=+2.0$^{\circ}$) is of unresolved depth ($<2$ kpc) and covers 45 square degrees of sky. At a distances of 19 kpc, the projected radius for this feature is 1.5 kpc. The 180$^{\circ}$ over-density was first seen in the \citet{Keller09_TNO} study (their Feature $C$) but without significant RRL population a distance was not able to be ascribed for the feature. 

\subsubsection{160$^{\circ}$ over-density}

The 160$^{\circ}$ over-density is centred at RA=160.9$^{\circ}$, Dec=13.3$^{\circ}$ and has an extent of 36 square degrees. It is located at a heliocentic distance of 17 kpc, at which distance the approximately circular feature presents a radius of 1.3 kpc. The 160$^{\circ}$ over-density is seen projected in front of the Sgr leading arm. In Figure \ref{figure:cmds} (top right) this Sgr material is apparent as the sub-giant population seen at $g_0 \sim 20.2$.

\subsubsection{The Virgo Equatorial Stream}

The Virgo Equatorial Stream (VES) is centred at (RA=203.3$^{\circ}$, Dec=+2.7$^{\circ}$, r=28 kpc) and covers at least 162 square degrees (since it is not clear that it is contained within the southern boundary of the SDSS field). It possess a small line-of-sight depth of $<$4 kpc. The corresponding projected dimensions of the VES at a mean distance of 28 kpc are 8.9$\times$4.4 kpc. 

The VES is situated slightly westward of the VOD but 10kpc more distant ($\Delta g = 1.6$ magnitudes).  Could the VES be a metal-rich sub-population of the VOD? The above derived distances assume an old (12Gyr), metal-poor population ([Fe/H]=-1.7). This is analogous to stellar population seen in present-day faint dwarf spheroidals in the local neighbourhood. If the observed structures formed from a much larger ($M>10^8M_{\odot}$) gas-rich system such as Sgr we might expect a broad metallicity distribution function for the debris. This is seen in the case of the Andromeda Giant Stellar Stream believed to have been recently derived from a $M\sim10^8M_{\odot}$ satellite \citep{Ibata07}. However, if we assume [Fe/H]=-1.7 for the VOD, the implied VES metallicity would be greatly super-solar. In addition, the work of \citet{An09} does not support the presence of a substantial spatial metallicity variation in this region. Similarly, it is not possible that the VES could result from the lower MS of the VOD since the VES is only 1.6 mag.\ more distant that the VOD and not the $\sim3$ mag.\ expected for the lower MS. For these reasons we conclude that the VOD and VES are separate spatial structures.

\section{Luminosities of sub-structure}

The number of sub-giants associated with each of the four features can be equated to an integrated absolute magnitude for the underlying stellar populations. For each pixel we form the sum of the number of stars in excess of the local background within our sub-giant box over the magnitude range of the feature. An old, metal-poor isochrone (as before, 12 Gyr, [Fe/H]=-1.7 \citet{Girardi04}) is then populated according to a Salpeter initial mass function (with masses$\ge 0.08 M_{\odot}$) until the observed number of sub-giants is reproduced \citep{Keller01}. We then form the integrated absolute magnitude of this population. Table \ref{table:mvs} reports the resulting absolute magnitudes for the features. 

The uncertainties in these values encompass the stochastic variation in multiple random realisations of the model population. There is also some variation in these luminosities depending on the assumed metallicity of the population. If we were to assume a more metal-rich population, say [Fe/H]=-0.4 such as seen in the core of Sgr, the derived luminosities would be 0.8 magnitudes fainter. Both the VOD and the Virgo Equatorial Stream extend outside the SDSS survey area. Consequently the absolute magnitudes we derive for these features represent lower bounds. In addition, in the case of the VOD we expect that the extent of the substructure along the line of sight will result in a further underestimation of its luminosity as discussed in Section \ref{section:VOD}.


For the VOD we derive an absolute magnitude of $M_{g}=-8.9$. \citet{Juric08} calculates an absolute magnitude of $M_{r}=-8.0$ mag.\ for the VOD from the sum of the excess stellar population between 18$<$$r$$<$21.5, a mean distance of 10 kpc, and an extent of 1000 square degrees. This equates to a surface brightness of 32.5 mag.\ arcsec$^{-2}$. As \citet{Juric08} state, this is a lower limit to the brightness of the substructure, since it does not include bright giants or the population below the magnitude limit ($r$=21.5). Furthermore, as we have discussed above, we claim that the \citet{Juric08} study overestimates the spatial extent of the VOD.

By evaluating the excess in observed luminosity functions to the VOD compared to the predictions of the Besancon Galaxy model, \citet{Prior09} calculate an absolute magnitude of $M_V$=-11.9 mag.\ (area of 760 square degrees and mean distance of 19kpc). The \citet{Prior09} study utilizes the luminosity function excess from $16.25<V<19.75$.

The present study differs in the determination of integrated luminosity from these previous studies since it does not sum over a magnitude range of main-sequence over-density. Rather, the number of excess subgiants measured here is the consequence of stellar evolution and an IMF thus the luminosities derived here are total luminosities for the over-densities in question.

\begin{table*}
\caption{The spatial and luminosity details of the sub-structures described in the present study. In the case of the VOD and Virgo Equatorial Stream the sky coverage (and hence the derived absolute luminosity) is a lower limit as the features are not enclosed in the SDSS area.}
\begin{center}
\begin{tabular}{ccccccc}
\hline
 Object &  RA & Dec & Sky &  Surface    &  Absolute & Mean\\
 Name  & Center & Center & Coverage &   Brightness &  Magnitude & Distance \\
 & (deg.) & (deg.) & (sq.\ deg.) & (mag./arcsec$^2$) & $M_g$ (mag.) & (kpc) \\
\hline
Virgo & & & & \\
Over-density  & 194.6 & 1.8 & 144 &  30.5 & $-8.9^{+0.2}_{-0.3}$ & 16\\\\
$160^{\circ}$ & 160.9 & 13.3 & 36 & 32.2 & $-5.6^{+0.3}_{-0.4}$ & 17\\\\
$180^{\circ}$ & 179.8 & 2.0 & 45 & 33.2 & $-5.1^{+0.4}_{-0.6}$ & 19\\
Virgo & & & & \\
Equatorial Stream & 203.3 & 2.7 & 162 & 31.7 & $-8.8^{+0.5}_{-0.7}$ & 28\\
\hline
\end{tabular}
\end{center}
\label{table:mvs}
\end{table*}

\section{Origins of sub-structures: Sgr debris?}
\label{section:origins}

The tenuous nature of the stellar populations of the VOD, 180$^{\circ}$, 160$^{\circ}$ and VES sub-structures suggests that they have originated from the tidal disruption of one or more progenitor galaxies. In the studies of \citet{Johnston08} and \citet{Font08} a comparison is made between simulated observable properties of a stellar halo and the parent galaxy's accretion history. The studies conclude the high surface brightness features ($\mu_{V} \le 30$ mag arcsec$^{-2}$) in the outer halo (R$>$10 kpc) usually originate from the accretion of single satellites. Furthermore, bright features are almost exclusively derived from relatively high mass progenitors (10$^7$-10$^9$ M$_{\odot}$). Since stellar evolution proceeds longer, and processed material is retained, in more massive galaxies, the material accreted is expected to be of moderate to high metallicity. In other words, the stellar populations of substructures are not expected to resemble old and metal-poor dwarf spheroidal satellites nor the underlying component of the stellar halo.

\citet{Johnston08} shows that these expectations are broadly consistent with our understanding of the outer halo of the Milky Way and M31. In the case of the Milky Way we find one surviving satellite of high surface brightness (Sgr) and it is composed of material significantly more metal-rich than the background halo population. Observations of M31 (see \citet{Ibata07} and references therein) reveal a halo dominated for R$>20$ kpc by the Giant Stellar Stream. The Giant Stellar Stream is comprised of relatively metal-rich debris and is thought to have been deposited  during the accretion of a high mass ($\sim10^{9}$ M$_{\odot}$) progenitor on an approximately radial orbit. The average surface brightness of the Giant Stellar Stream is $\mu_{V} = 30\pm0.5$ mag\ arcsec$^{-2}$ \citep{Ibata01}. \citet{Ibata07} note six other substructures, some of which possess a metallicity distribution function similar to the Giant Stellar Stream and are potentially related to the same accretion event. Morphologically these substructures appear as streams, arcs, shells and irregular structures superimposed on a metal-poor halo.

Given the apparent dominance of Sgr debris in outer halo substructure, and the finding from cosmological simulations that the probability of additional high mass accretion events is low for the Milky Way, can the observed substructure of the present study be explained by Sgr debris?

\begin{figure}[h]
\begin{center}
\includegraphics[scale=0.55, angle=0]{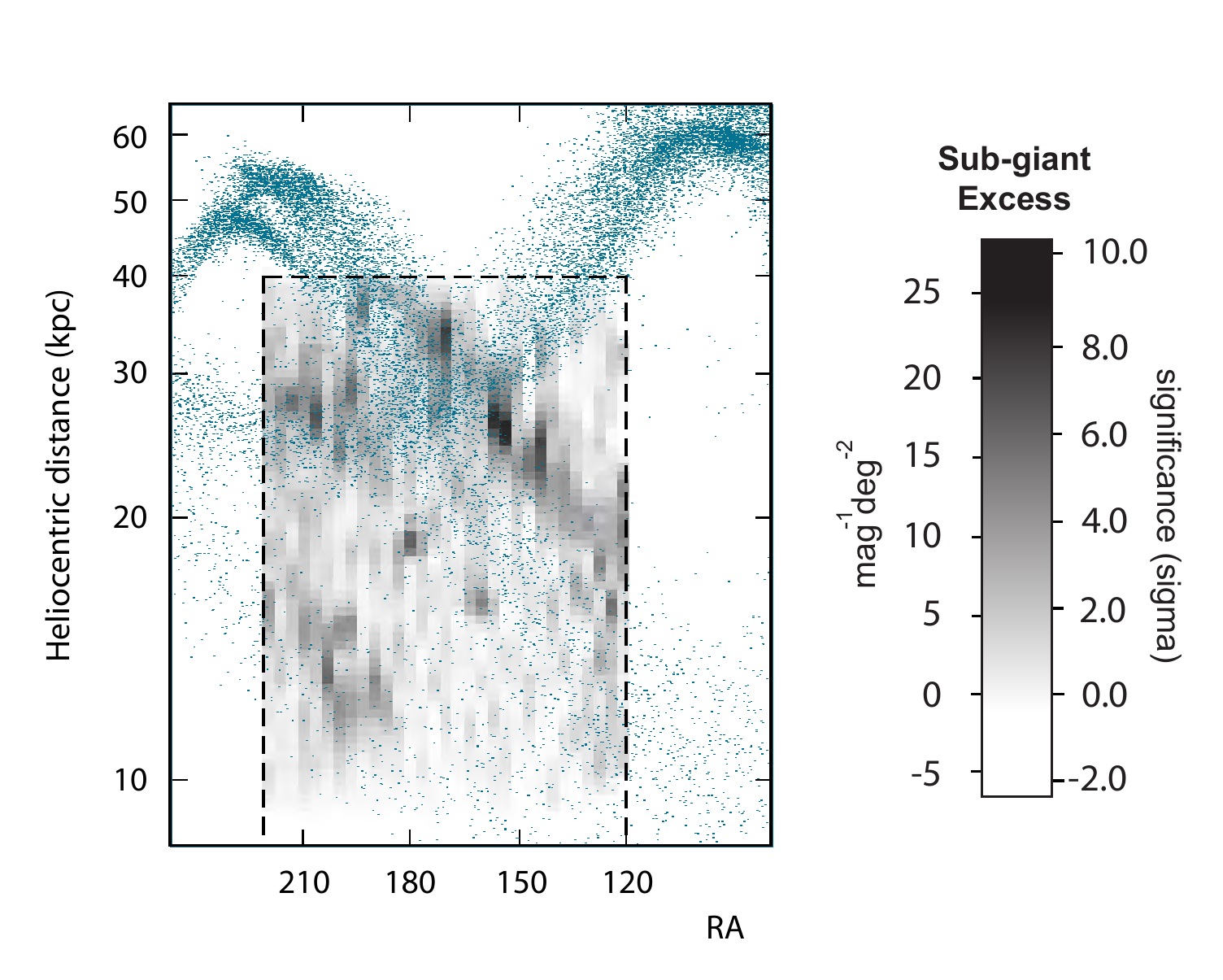}
\caption{We compare the Sgr model of \citet[][prolate model that best fits the leading arm material]{Law05} (points) with our observations of halo sub-giant excess (greyscale, our survey limits are shown by the dashed lines). The discordance between observed and simulated leading arm material is apparent at RA$<$170$^{\circ}$ where the observed leading arm is at greater distances than expected from the Sgr model.}\label{figure:Sgr}
\end{center}
\end{figure}

In Figure \ref{figure:Sgr} we compare the best-fitting prolate model ($q$=1.25) from the N-body simulations of \citet{Law05} to the over-densities of the present study. As we have seen in Section \ref{section:method} the choice of metallicity does significantly modify the derived distance to the material and here we assume a single global metallicity of [Fe/H]=-1.7 that is significantly less than the mean metallicity of the leading arm material \citep[Fe/H=-1.1]{Chou07}. It is also the case that there is a strong metallicity gradient along the leading arm material from the core of [Fe/H]=-0.4 \citep{Chou07}. These caveats acknowledged, some qualitative conclusions can be made between the distribution of model debris particles and the observed sub-giant excess in the Sgr stream. 

A clear discrepancy is seen in the model's inability to match the Sgr leading arm: namely that for RA$<$170$^{\circ}$ the Sgr leading arm material is at a greater distance at given RA than the model predicts. The stellar over-density may be brought into line with the model distribution by assuming a much higher metallicity for the debris ([Fe/H]$\geq-0.4$). More problematically, the metallicity distribution function for the debris would have to be increasingly metal-rich as we proceed {\it{away}} from Sgr, a conclusion discrepant with existing measurements by \citet{Chou07}.

To date no one N-body simulation has been capable of satisfactorily reproducing all of the extant observational data \citep{Law09}. The angular precession of the leading arm indicates a spherical or slightly oblate halo \citep{Fellhauer06,MartinezDelgado07} whereas, the distance to, and radial velocity of, the leading arm material indicates a prolate halo \citep{Helmi04}. These conclusions have been drawn from simulations that are axisymmetric. The recent triaxial models of \citet{Law09} are able to replicate the data, however as pointed out by \citet{Law09} these models are somewhat unsatisfactory since they are in what is expected to be an unstable configuration. The observations of \citet{Prior09b} mirror the conclusions of \citet{Law05} that the current models of Sgr disruption are inadequate as they do not consider the orbital evolution of Sgr. Furthermore, as demonstrated by many simulations (see for example, \citet{Bekki08}) it may be necessary to consider the evolution of both the luminous and dark matter components of the Sgr dwarf as such treatment modifies orbital evolution significantly.

Clearly additional, and more complex, modelling is vital to clarify the Sgr debris streams. Given the large discrepancy between the current models and our observations it is premature to make any definitive remarks about the association of this study's sub-structure to the Sgr debris stream. Currently, it is possible that the Virgo Equatorial Stream could be associated with the younger trailing arm material (see Figure \ref{figure:Sgr}) and that the VOD, 180$^{\circ}$ and 164$^{\circ}$ sub-structures may represent over-densities along the older trailing arm. However, it must be acknowledged that such possible associations may not be upheld by radial velocity determinations and revised modelling. 

In addition to critically needed modelling, further sky coverage in the southern hemisphere is required to trace the spatial extent of these substructures. This will soon be provided by the SkyMapper telescope's Southern Sky Survey \citep{Keller07}. Radial velocities and metallicities for samples in the sub-structures revealed by this study are needed. They will enable us to discern if the features can be accommodated by revised Sgr models or if they represent the remnants of distinct merger events. 

\section{Summary}

An examination of the excess of sub-giant stars over a subset of the SDSS DR6 reveals a series of extended halo over-densities in the survey heliocentric distance range of 10 - 40 kpc. The most prominent feature is that of the Sagittarius dwarf galaxy (Sgr) leading tidal debris arm. We derive a spatial configuration for the Sgr leading arm material that is in agreement with previous studies. The Virgo Over-density is seen to extend from 12-18 kpc, with a mean distance of 16 kpc, and to cover an area of 144 square degrees. This is considerably smaller extent than found in previous studies. We contend this is due to our resolution of the region into a series of sub-structures. We determine the surface brightness of the VOD to be 30.5 mag/arcsec$^{2}$. 

Three new candidate halo substructures are detected. They consist of two roughly spherical over-densities, the 160$^{\circ}$ and 180$^{\circ}$ features. These features lie at 17 and 19 kpc respectively and present a surface brightness of 32.3 and 33.2 mag/arcsec$^{2}$ respectively. The third feature, the Virgo Equatorial Stream, lies at 28 kpc and extends over at least 162 deg$^{2}$. It possesses a stream-like morphology and a surface brightness of 31.7 mag/arcsec$^{2}$. 

We compare our observations of halo substructure to the predictions of Sgr tidal debris. We highlight a critical deficiency of current modelling of Sgr, namely, the discordance between the observations and the model predictions for the distance to the leading arm material. We conclude that further modelling effort and radial velocities are required to clarify the origins of the substructures observed.

\section*{Acknowledgments}

The author wishes to thank Prof.\ Gary S.\ Da Costa for discussions and comments on the draft of this paper. This research has been supported in part by the Australian Research Council through Discovery Project Grants DP0343962 and DP0878137.

    Funding for the SDSS and SDSS-II has been provided by the Alfred P. Sloan Foundation, the Participating Institutions, the National Science Foundation, the U.S. Department of Energy, the National Aeronautics and Space Administration, the Japanese Monbukagakusho, the Max Planck Society, and the Higher Education Funding Council for England. The SDSS Web Site is http://www.sdss.org/.

    The SDSS is managed by the Astrophysical Research Consortium for the Participating Institutions. The Participating Institutions are the American Museum of Natural History, Astrophysical Institute Potsdam, University of Basel, University of Cambridge, Case Western Reserve University, University of Chicago, Drexel University, Fermilab, the Institute for Advanced Study, the Japan Participation Group, Johns Hopkins University, the Joint Institute for Nuclear Astrophysics, the Kavli Institute for Particle Astrophysics and Cosmology, the Korean Scientist Group, the Chinese Academy of Sciences (LAMOST), Los Alamos National Laboratory, the Max-Planck-Institute for Astronomy (MPIA), the Max-Planck-Institute for Astrophysics (MPA), New Mexico State University, Ohio State University, University of Pittsburgh, University of Portsmouth, Princeton University, the United States Naval Observatory, and the University of Washington.


\end{document}